\documentclass[10pt, conference, compsocconf]{IEEEtran}
\usepackage{cite}
\usepackage[hidelinks]{hyperref}

\usepackage[pdftex]{graphicx}
\usepackage[cmex10]{amsmath}
\usepackage{algorithmic}

\usepackage{url}
\usepackage{amssymb}
\begin{document}
\title{Unmasking Communication Partners: A Low-Cost AI Solution for Digitally Removing Head-Mounted Displays in VR-Based Telepresence}

\author{\IEEEauthorblockN{Philipp Ladwig\IEEEauthorrefmark{1},
Alexander Pech\IEEEauthorrefmark{1},
Ralf Dörner\IEEEauthorrefmark{2} and
Christian Geiger\IEEEauthorrefmark{1}}
\IEEEauthorblockA{\IEEEauthorrefmark{1}
University of Applied Sciences Düsseldorf, Germany\\
Mixed Reality and Visualization Group (MIREVI)\\
\{philipp.ladwig@, alexander.pech@study., geiger@\}hs-duesseldorf.de}
\IEEEauthorblockA{\IEEEauthorrefmark{2}RheinMain University of Applied Sciences\\
Faculty of Design – Computer Science – Media, Wiesbaden, Germany\\
ralf.doerner@hs-rm.de}
}
\maketitle
\begin{abstract}
Face-to-face conversation in Virtual Reality (VR) is a challenge when participants wear head-mounted displays (HMD). A significant portion of a participant’s face is hidden and facial expressions are difficult to perceive. Past research has shown that high-fidelity face reconstruction with personal avatars in VR is possible under laboratory conditions with high-cost hardware. In this paper, we propose one of the first low-cost systems for this task which uses only open source, free software and affordable hardware. Our approach is to track the user's face underneath the HMD utilizing a Convolutional Neural Network (CNN) and generate corresponding expressions with Generative Adversarial Networks (GAN) for producing RGBD images of the person's face. We use commodity hardware with low-cost extensions such as 3D-printed mounts and miniature cameras. Our approach learns end-to-end without manual intervention, runs in real time, and can be trained and executed on an ordinary gaming computer. We report evaluation results showing that our low-cost system does not achieve the same fidelity of research prototypes using high-end hardware and closed source software, but it is capable of creating individual facial avatars with person-specific characteristics in movements and expressions.
\end{abstract}
% One solution is to employ artificial intelligence (AI) to create images of the participants’ face where the HMD is digitally removed. 

% \fi
\IEEEpeerreviewmaketitle

\section{Introduction}
%VR promises great potential in a variety of applications. Especially low-cost VR that employs head-mounted displays is becoming more commonplace. However, authentic social interaction in VR is still an unsolved problem especially if faces are hidden behind an HMD and it is only feasible to represent communication partners as stylized avatars. But particularly the current situation due to the global COVID-19 pandemic shows that the need for authentic social participation has probably never been greater. Even with video conferencing replacing classic voice-based telephone conversations, it soon became clear that this is no satisfactory substitute for real-life meetings. However, as technology is constantly evolving, the next logical step after adding 2D video to telephony is to add a third dimension and increase presence to obtain telepresence. 

Several science fiction writers have already enthusiastically envisioned immersive technologies that allow us to teleport our avatars to distant places and act as if we were actually present. Today it is technically feasible to create digital avatars and animate them in real time in virtual environments. However, creating a digital avatar is usually associated with a lot of work during recording and post-processing. Especially the modeling of the face requires a skilled 3D artist to avoid the Uncanny Valley effect \cite{UncannyEffectAusFidler07} as even tiny discrepancies between a real and an artificially modeled face are often easily spotted and may lead to an unpleasant or even eerie impression. That is one reason why the majority of current applications for social VR deliberately avoid a realistic appearance of avatars. However, recent advances in deep neural networks show promising results for authentic avatar representations in virtual worlds, which are often trained end-to-end and bridge the Uncanny Valley successfully\,\cite{Tewari2020NeuralSTAR}.  

While rapid avatar creation and animation have been studied over decades \cite{STAR-3DMM}, the research field of expressive VR telepresence in combination with face tracking under an HMD is an emerging topic. It still poses challenges such as face tracking from extreme camera positions and  time-consuming as well as expensive capture procedures\,\cite{Lombardi18DAM, Wei19, ModularCodecAvatars20, richard2020audio}. Especially the feasibility of using low-cost and commodity hardware for expressive VR telepresence is rarely addressed, and appropriate solutions have not been made publicly available. Instead of aiming for utmost visual quality, we strive for a solution that can be employed in real-world applications without prohibitive costs. Our contributions are:

% [Vor Our contributions are]: Thus, we focus on utilizing commodity and low-cost hardware and explore to what extent our results are acceptable for users and how they compare to the visual quality achieved by other systems\,\cite{Lombardi18DAM, Wei19}. 

\begin{itemize}
    \item an end-to-end learning GAN and an automatic procedure for capturing and reconstructing 3D faces with individual expressions;
    \item a CNN and a procedure for facial landmark tracking of the face while the user wears an HMD;
    \item the combination of above items into a system for capture, reconstruction, and real-time animation of individual facial avatars for VR telepresence;
    \item a working prototype of our system as a proof of concept where we freely provide parts of the source code;
    % \item an assessment of the visual quality achievable with our system;
\end{itemize}

% The structure of the paper is as follows:
% - x \\
% - y \\
% - z \\ 
% \\

\section{Related Research}
% \subsection{Face Reconstruction and Tracking with HMD}
Face reconstruction and tracking without an HMD is a well-researched topic. A detailed state of the art report can be found in \cite{starMonoFace18}. In contrast, tracking the users’ expressions while they wear an HMD is a challenging task and a new area of research. Li et al.\,\cite{Li15FacialHMD} attached an RGBD camera with an outreaching arm to an HMD that tracks expressions around the mouth. Additionally, strain sensors in the foam liner of the HMD recorded motion around the eyes. Casas et al.\,\cite{Casas16Rapid} extended the system from Li et al.\,\cite{Li15FacialHMD} with the use of personal 3D scans of persons and blendshape animation between expressions. The optical results of this work is similar to ours, but we use a holistic neural representation as a textured depth map instead of a set of textured meshes with linear blendshape interpolation. Olszewski et al.\,\cite{Olszewski16HighFid} used an RGB camera attached to the HMD along with a CNN to create plausible mouth animation. They realized a direct regression from images to blendshape parameters of their generic face model. This approach requires a significant amount of manual work of 3D artists, which have do define authentic blendshapes for the avatar. Compared to our work, we propose an end-to-end solution for learning an personal avatar without manual intervention and our system automatically regress facial expressions to head-mounted camera (HMC) images. Nevertheless, our approach does not achieve the same visual quality of mouth tracking. While Li et al.\,\cite{Li15FacialHMD} and Olszewski et al.\,\cite{Olszewski16HighFid} used generic head models, Thies et al.\,\cite{Thies18FaceVR} used an personal image-based avatar based on an analysis-by-synthesis approach.

Well known methods such as 3D Morphable Models (3DMM)\,\cite{3dmm} offers identity preserving avatars, however, they usually fail to bridge the Uncanny Valley. The animation of eyes, oral cavity or highly-deformable areas of the face often lack natural appearance due to changes in specularity, blood flow or subsurface scattering effects. Traditional solutions use light transport formulation which tries to recreate natural appearance which are usually computation-intensive tasks. However, the usage of deep neural networks enables a new data-driven pipeline and is capable of capture and reconstruct complex nonlinear effects. Therefore, our approach is different to the idea behind 3DMMs, because we do not use blendshapes or complex light transport formulations. Our model is stored as a compact latent representation of a real face in which many effects, such as complex changes of materials due to deformation, were handled naturally.

The most related systems to ours were proposed by Lombardi et al.\,\cite{Lombardi18DAM}, Wei et al.\,\cite{Wei19}, Chu et al.\,\cite{ModularCodecAvatars20} and Richard et al.\,\cite{richard2020audio}. They used a multi-view capture apparatus with 40 cameras pointed at the front hemisphere of the face to create a detailed facial model called Deep Appearance Model. This model is trained with a conditional variational autoencoder (VAE) and is capable of generating a view-conditioned texture and geometry for a high-fidelity facial avatar. The capture process, post processing and application requires high-cost laboratory equipment and processing power. It is closed source software and can not be used with commodity hardware or trained and executed on a computer with a single GPU, which is a difference to our system. Another difference is that Deep Appearance Model consists of a triangle mesh and employ blendshapes while we render time-varying textured point clouds without triangulation. Another difference is the way of driving the avatar's expressions. The works of Lombardi et al.\,\cite{Lombardi18DAM}, Wei et al.\,\cite{Wei19} and Chu et al.\,\cite{ModularCodecAvatars20} regress HMC images to expression parameters of the Deep Appearance Model. Each work uses slightly different ways to bridge the domain gap between HMC images and the face representation in order to find their correspondence with the help of a differentiable rendering approach. However, none of these works use facial landmarks directly. As most sophisticated system Wei et al. uses the CycleGAN by Zhu et al.\,\cite{UnpairedI2I-Zhu17} to realize an unpaired image-to-image translation between real images and rendered images. Our proposed system skips the differentiable rendering stage and we train our GAN to directly generate 3D output as RGBD images from landmark. Instead of establishing a mapping between 2D landmarks to 3D vertices of a mesh, which is often difficult, we employ an image-to-image translation approach. This allows for a significantly less complex application structure, training time, and inference time, requires less-detailed input data of the persons' face, and requires only commodity and low-cost hardware at the expense of the final reconstruction quality of the avatar.

\section{Design Rationale}
Our goal is to create a system that uses open source and free software in combination with low-cost hardware for automatic capturing, creating, and driving a 3D face avatar in real time for VR telepresence. 'Automatic' means that no manual modeling, adjustment, or landmark editing is required. The avatar should convey the identity of the person with the ability to transmit individual-typical expressions. Therefore, the system must allow for capture and reconstruction of personal expressions without relying on generic expression templates. A data-agnostic approach is intended in such a way that theoretically any RGBD sensor such as Microsoft Kinect, Intel RealSense, or depth sensors of smartphones could be used for capture.
% Past research has shown the quality and possibilities of a parametric model such as 3DMM\,\cite{3dmm}. However, in the field of our application, their main drawbacks are, first, the lack of free software for creating 3DMMs with eyes and oral cavity and, second, the inferior visual quality compared to the results of the state-of-the-art neural network architectures\,\cite{Tewari2020NeuralSTAR}. Though recent works combine the advantages of 3DMMs and neural networks, these works are also not publicly available or are complex in usage due the lack of documentation. An excellent overview of the state of the art regarding 3DMMs in combination with improvements via neural nets is presented by Egger et al.\,\cite{STAR-3DMM}. 
%The state of the art report on Neural Rendering of Tewari et al.\,\cite{Tewari2020NeuralSTAR} highlights and summarizes the capabilities of GANs for image synthesis. As the images generated by GANs are often hardly distinguishable from real images, Neural Rendering is considered an alternative to the classic rendering techniques rasterization and ray tracing. The ability of GANs to produce faces that successfully bridge the gap of the Uncanny Valley\,\cite{Mori-UncannyValley} and allow for end-to-end learning of an image-to-image translation task without manual adjustment make them particularly suitable for our application.

A further important requirement of the proposed system are real-time capabilities for consumer-grade computers with one GPU. Related works achieve high visual quality, but they use multiple GPUs for encoding and decoding\,\cite{Wei19, Lombardi18DAM, ModularCodecAvatars20}. Therefore, our system uses depth maps instead of meshes or voxels because the data structure of depth maps is better suited for fast processing on neural nets due to the fact that they only employ two-dimensional arrays instead of three-dimensional data. In addition, our CNN for face tracking is created with the intention to be lightweight and can be used simultaneously with the generator of our GAN on a single consumer-grade GPU with reasonable real-time speed. 

For an immersive telepresence application, it would be sufficient to reconstruct only the face area of the user, since non-hidden areas could be recorded and transmitted by additional external RGBD cameras in a room. Therefore a reconstruction of the entire head is not necessary and the reconstructed face could be merged into a single point cloud consisting of streams from multiple sensors distributed in the room.

\section{System Overview}\label{sec:systemOverview}
% \begin{figure*}
% \centerline{\includegraphics[width=\textwidth]{img/banner_besser.png}}
% \caption{Example of a figure caption.}
% \label{fig:banner}
% \end{figure*}
Creation and animation of a personal face avatar with our system requires only a few steps and a minimal amount of time and manual effort. Capturing the persons face with expressions requires 15min and additional 19\,hours for training. After training, the generator module of our GAN can be used for real-time interference in VR. The system works completely automatically. 
% It is only necessary to put on a helmet mount, performing a set of expressions and adjust 2 parameters for eye brow tracking. No further adjustments are necessary.

To give an overview of our system, we list 5 consecutive steps requiring for creating and driving a personal face avatar: 1.) The user`s face is captured with the help of a helmet mount (section \ref{sec:dataset}) as an RGBD image data set (requires 10\,min) and as an RGB image data set for the lower-face tracking (requires 5\,min).  2.) The data set is preprocessed for training, which involves clipping the facial area and creating corresponding facial landmark maps (FLM) for each RGBD image (Fig.\,\ref{fig:inference}, right column). 3.) The proposed 'RGBD-face-avatar GAN' is trained using the RGBD data set with the according FLMs from step 2 in order to learn an image-to-image correspondence. Furthermore, our lower-face tracking CNN is trained with the RGB data set.  4.) After training, the user can put on the HMD, and his/her facial expressions are captured by the HMCs. Each HMC covers only a small portion of the user`s face, and the image data of each HMC is used to generate a combined FLM 30 times per second. 5.) These FLMs are used as input for inference of our RGBD-face-avatar GAN which allows for generating the corresponding expression of the user's avatar face as RGBD images in real time.

The source code, videos of the system and additional materials can be found at \url{https://github.com/Mirevi/UCP-Framework}.
% Along with the FLM, the positional tracking data of the HMD is used for rendering the reconstructed face to the according tracked position of the user's head.

\section{Data Set}\label{sec:dataset}
% The following section describes the capture process, data handling as well as preparation for the training of the RGBD face avatar GAN and the inference stage.

\begin{figure}[b]
\centerline{\includegraphics[width=\linewidth]{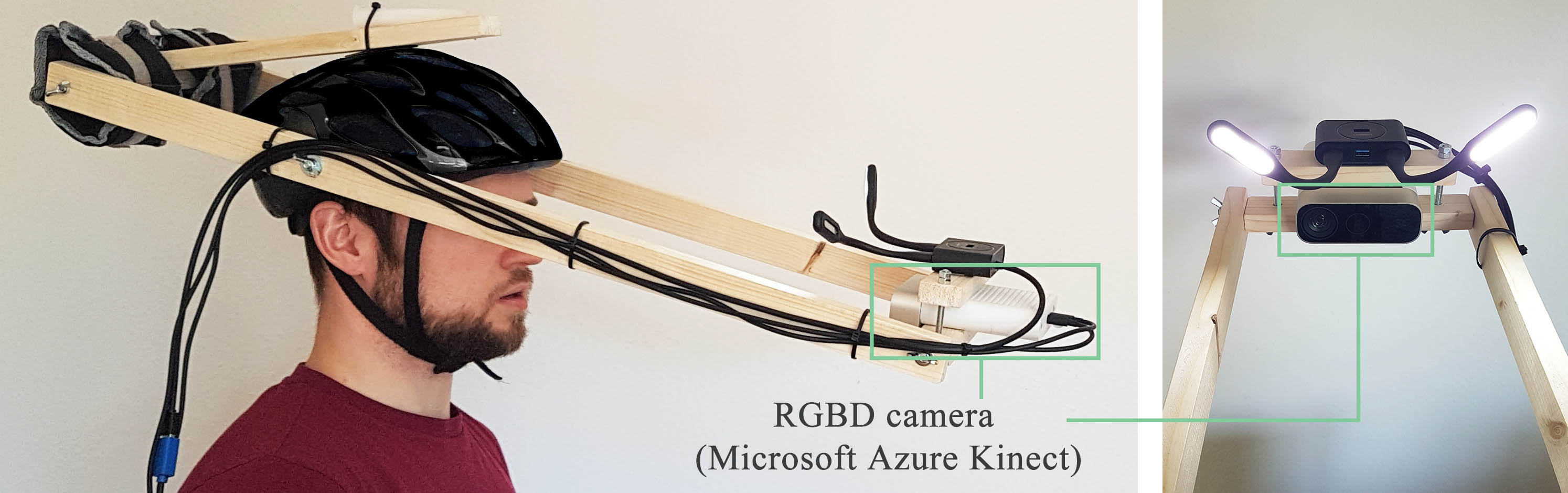}}
\caption{The helmet mount captures the face from a static distance without any tilting or rotation. The price for materials of the mount is less than 60\,USD without the Microsoft Azure Kinect.}
\label{fig:helmet}
\end{figure}

In order to create a person-specific face avatar we record a data set of around 600 RGBD images for facial reconstruction with according infrared (IR) raw images of the depth sensor for eye tracking. This amount was proven to be a reasonable tradeoff between recording time, training time and the resulting reconstruction results. It comprises 26 facial expressions (captured several times) and speaking 20 phonetically balanced sentences as well as about 5 minutes of talking.

It is beneficial for the reconstruction quality and the amount of time required for capture and training of a GAN if the training data contains only information that is relevant for the reconstruction. In our scenario, such information does not include data about different distances between the sensor and the face as well as tilting or any other rotation of the face because the movement of the avatar face will be controlled by the tracked pose of the HMD later. Therefore, we constructed a mount for a helmet with an attached RGBD sensor, as shown in Fig. \ref{fig:helmet}. The mount allows for recording solely frontal images without any face rotation from a static distance relative to the face. 

% The RGBD images contain 8-bit resolution per channel. Although the native resolution of the depth channel of many sensors is usually much higher (e.g. 16-bit for the Microsoft Azure Kinect and Intel RealSense), only a fraction of this range is needed for facial reconstruction. Since the distance between sensor and face remains the same in our mount during capture, we clip depth values in front of and behind the face. The depth values are reported in millimeters by the sensors, hence the 8-bit depth resolution leads to a range of 254\,millimeters which is sufficient for recording the depth of facial movements. As a side effect, the reduced depth range accelerates training and inference compared to using 16-bit resolution.

\begin{figure}[t]
\centerline{\includegraphics[width=6cm]{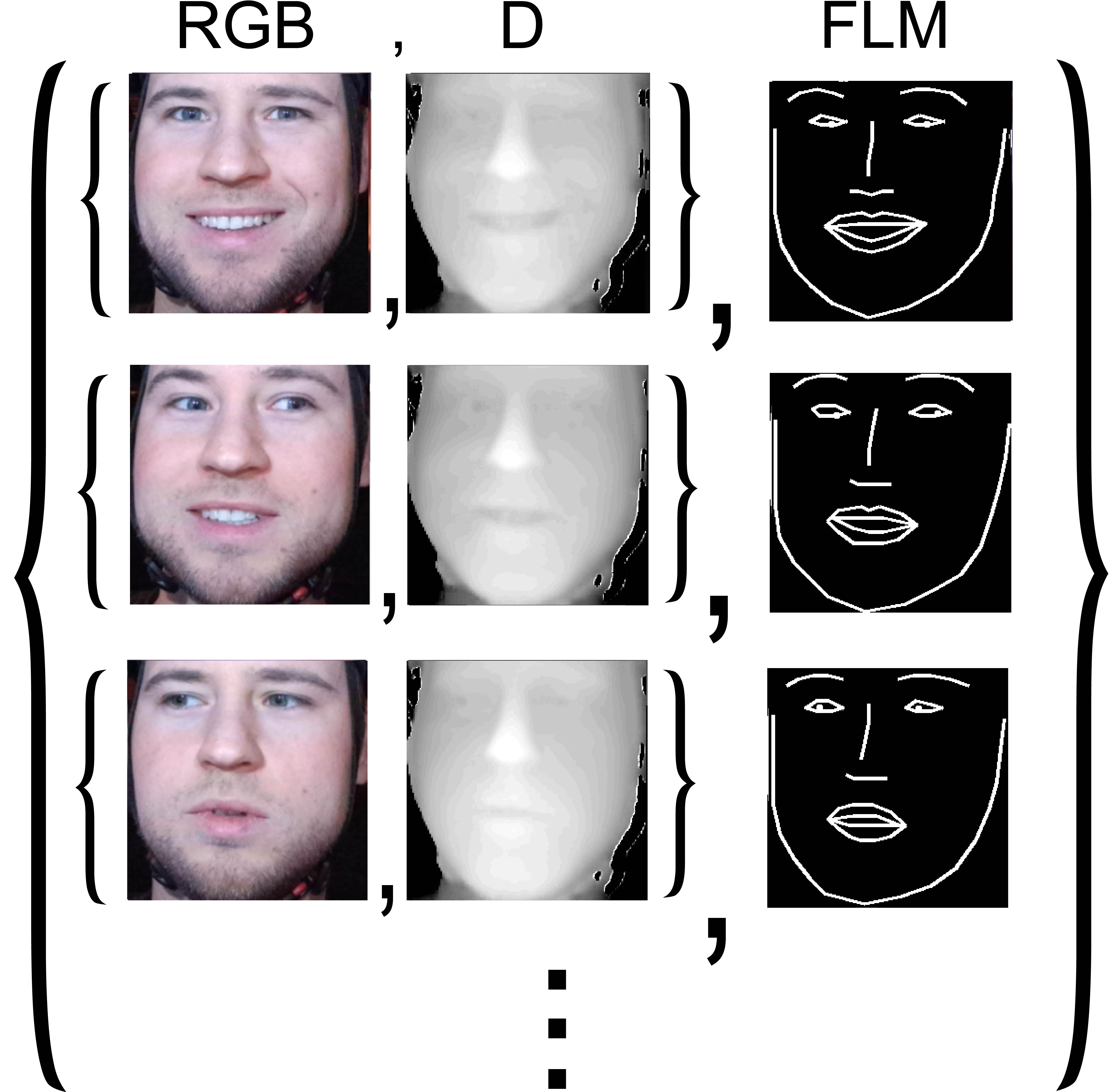}}
\caption{Paired image data set: For every pair of RGBD image made with the helmet mount a Facial Landmark Map (FLM) is created. During training, the RGBD-face-avatar GAN learns the paired image-to-image mapping from an FLM to an RGBD image.}
\label{fig:inference}
\end{figure}

In order to be able to control the output of the RGBD-face-avatar GAN at inference, a mapping between given input and output images must be learned. Therefore, the GAN receives a training set of aligned image pairs (Fig.\,\ref{fig:inference}).
These pairs consist of RGBD images, which were captured by the RGBD sensor in the helmet mount (Fig.\,\ref{fig:helmet}), and the FLMs, which are binary images containing 70 facial landmarks (Fig.\,\ref{fig:inference}, right column). These facial landmarks are automatically generated based on the RGB and infrared images provided by the sensor in the helmet mount and allow for saving the expression of the user in an abstract form. This abstract form of a FLM can later be generated by our face tracking HMD and serves as input for the generator of the RGBD-face-avatar GAN at inference.
\begin{figure*}[h]
\centerline{\includegraphics[width=\textwidth]{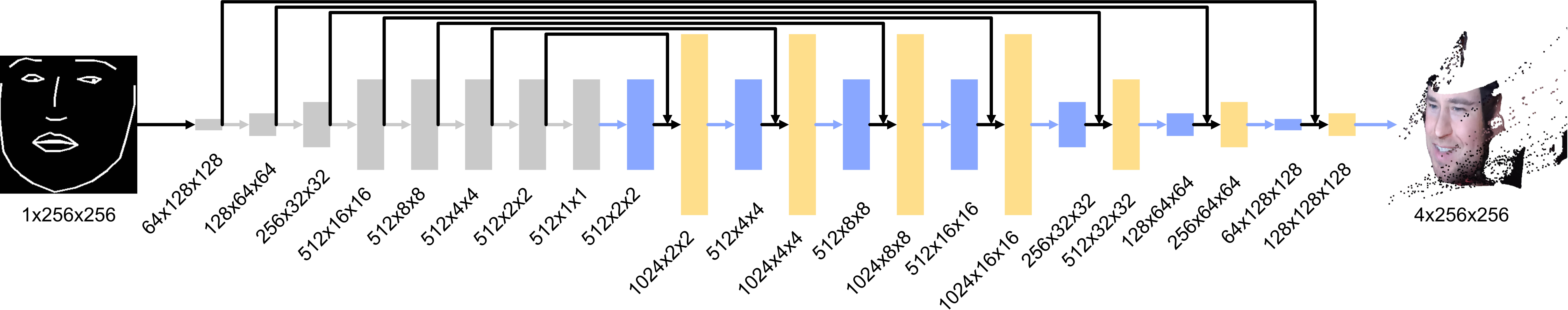}}
\caption{Generator of the RGBD-face-avatar GAN.}
\label{fig:Generator}
\end{figure*}
\begin{figure}[t]
\centerline{\includegraphics[width=7.0cm]{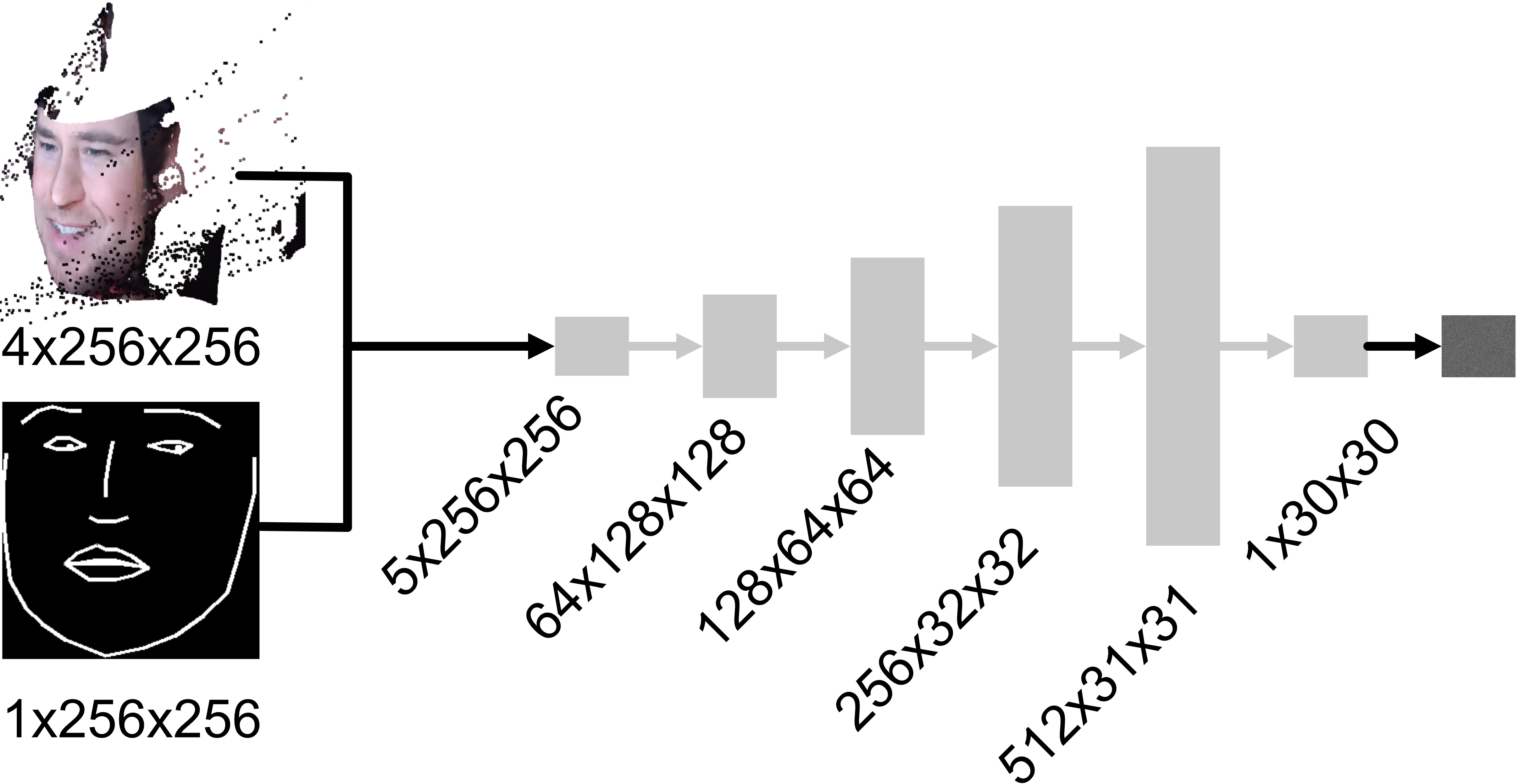}}
\caption{Discriminator of the RGBD-face-avatar GAN.}
\label{fig:Discriminator}
\end{figure}
The automatic creation of the landmarks in the training data set is realized by the openly available Facial Alignment Network (FAN) by Bulat and Tzimiropoulos\,\cite{bulat2017-FAN-HowFarAreWe}. It is a CNN that provides image coordinates of 68 landmarks based on a given RGB face image. Since the FAN does not provide landmarks for the gaze direction or iris position, we implemented our own eye tracking algorithm. Therefore, we use the infrared images provided by the time-of-flight sensor in the mount because gaze estimation on infrared images is more reliable than on RGB images\,\cite{Ir-EyeGaze-Works}. We concatenate the gaze information to the FLM as two additional landmarks. Eventually, the training set consists of aligned pairs of RGBD and corresponding images with the expression as FLMs.

\section{Architecture of the RGBD-face-avatar GAN}
We built our RGBD-face-avatar GAN upon the Pix2Pix GAN by Isola et al.\,\cite{pix2pix2016}. Compared to the original version, we added a fourth feature map to the generators output in order to be able to output a depth channel. As further addition, the discriminator receives five feature maps instead of six. While the first four feature maps correspond to the channels of an RGBD image, the remaining one contains the corresponding FLM. 
% The generator of our GAN~$\mathcal{G}_\phi$ at time instant $t$.
% \begin{equation}
%     (T_t, D_t) = \mathcal{G}_\phi(l_t)
%     \label{eq:gan_function}
% \end{equation}
% The input vector~$l$ contains the facial landmarks. The output of $\mathcal{G}_\phi$ is a tuple $(T, D)$ that contains a RGB texture $T\in \mathbb{R}^2$ and a depth map $D\in \mathbb{R}^2$. 
% $l \in \mathbb{R}^N$
Let \(\mathbf{L}\in \mathbb{R}^{2\times 70}\) be the FLM, which contains 70 facial landmarks in image coordinates, we can illustrate the procedure as
\[ \mathbf{T}_t, \mathbf{D}_t \leftarrow \mathcal{G}\phi(\mathbf{L}_t), \]
where \( \mathcal{G}\phi \) is the generator of our RGBD-face-avatar GAN which produces \( \mathbf{T} \in \mathbb{R}^{256x256} \) as an RGB texture and \( \mathbf{D} \in \mathbb{R}^{256\times 256} \) as depth map at time instant \(t \). A rendered image \( \mathbf{R} \in \mathbb{R}^{w\times h} \) of the face can be rendered from a rasterizer \(\mathcal{R}\):
\[ \mathbf{R}_t \leftarrow \mathcal{R}(\mathbf{T}_t, \mathbf{D}_t, \mathbf{C}_t),\]
where \(\mathbf{C} \) denotes the camera position and  projection function.

We trained the network for over 200 epochs with a batch size of 1 with random jitter and without mirroring. The learning rate was 0.0002 for the first 100 epochs and decreased linearly for the remaining 100. Weights were initialized from Gaussian distribution with a mean of 0 and a standard deviation of 0.02. The architecture of the discriminator and the generator are shown in Fig.\,\ref{fig:Generator} and Fig.\,\ref{fig:Discriminator}.

\section{Face Tracking HMD and Reconstruction}
\begin{figure}[b]
\centerline{\includegraphics[width=\linewidth]{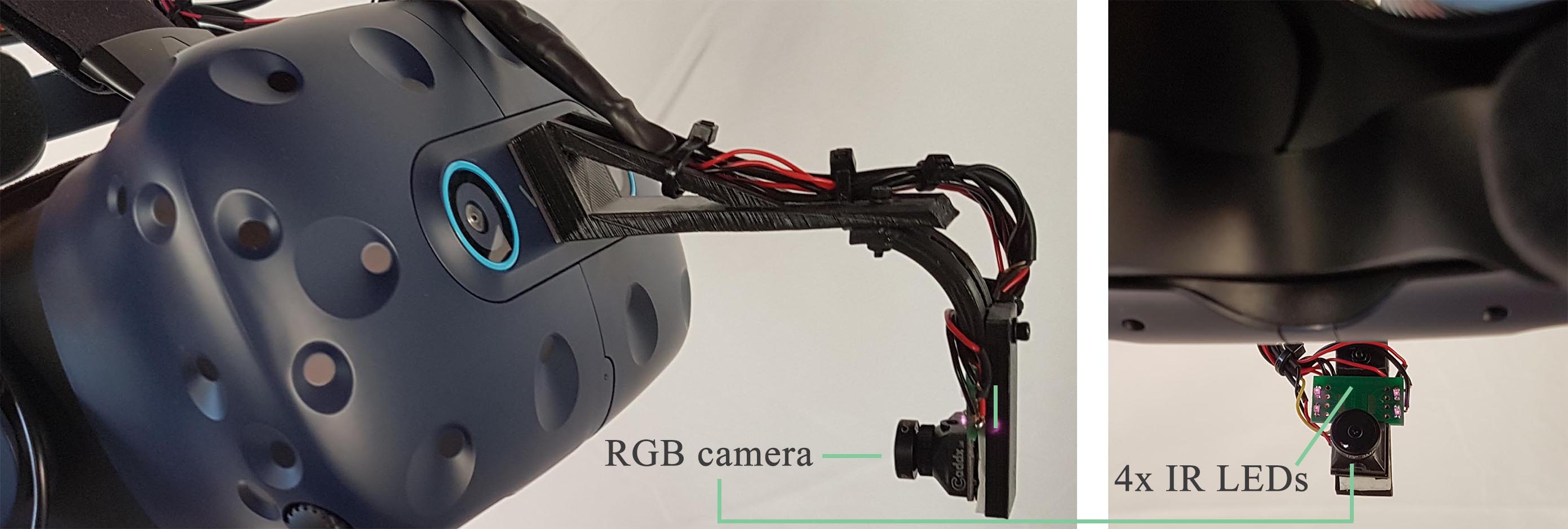}}
\caption{Face tracking HMD with 3D-printed mount for camera and IR LEDs.}
\label{fig:Face-Tracking-Vive}
\end{figure}
For capturing the expressions of the user while wearing an HMD and for translating these tracking data to the corresponding expressions of the avatar, the image streams of 5 HMCs that are attached to the HMD are processed. We use the HTC Vive Pro Eye HMD\,\cite{ViveEye}, which is, at delivery, already equipped with two cameras behind the Fresnel lenses for eye tracking. In combination with the SRanipal runtime of Vive\,\cite{SRanipal}, these cameras deliver data on eye gaze direction and eye openness and are concatenated into the FLM.

In addition to the eye tracking cameras of the HMD we added another 3 miniature cameras with IR filter. Two of those cameras are used for tracking the eyebrows and another one for tracking the lower face area. The cost for one camera and a video-to-USB converter is approximately 40\,USD. The following two subsections describe the data processing and landmark extraction in detail.

\subsection{Lower-face tracking CNN}\label{sec:lower-faceTrackingCNN}
\begin{figure}[h]
\centerline{\includegraphics[width=\linewidth]{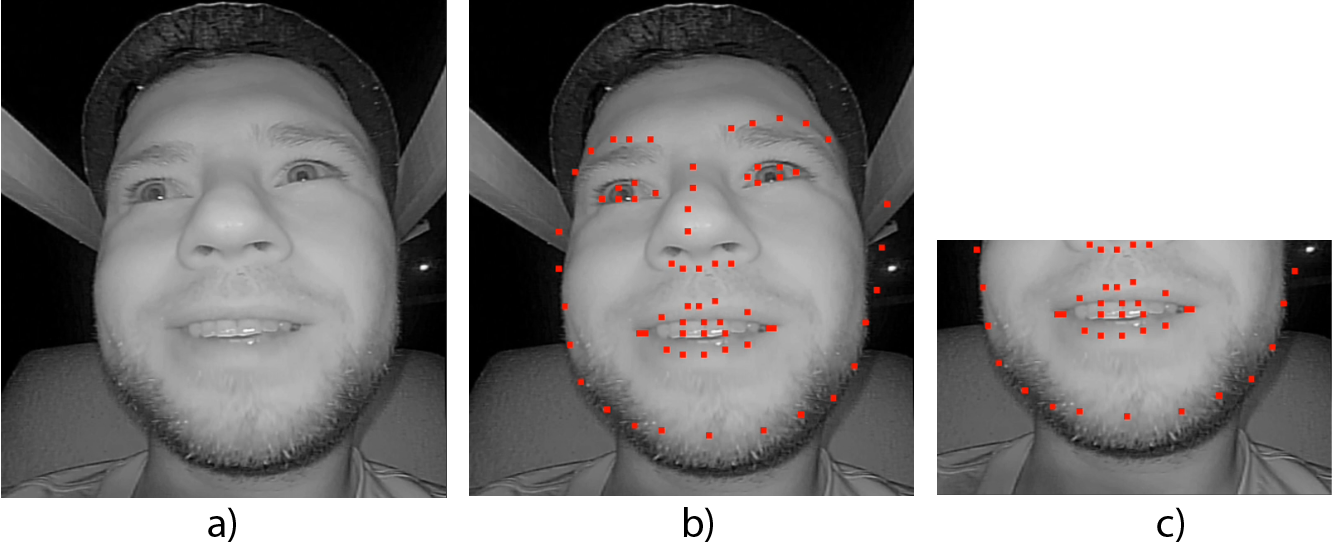}}
\caption{Creation of the training set for the lower-face tracking CNN. a) shows the uncovered face captured from same angle to the one in the face tracking HMD. b) The Facial Alignment Network\,\cite{bulat2017-FAN-HowFarAreWe} determines 68 landmarks. c) The upper face is cropped and the remaining image area and landmarks are used as labeled training set for the lower-face tracking CNN.}
\label{fig:cnnTraning}
\end{figure}
\begin{figure}[b]
\centerline{\includegraphics[width=7.0cm]{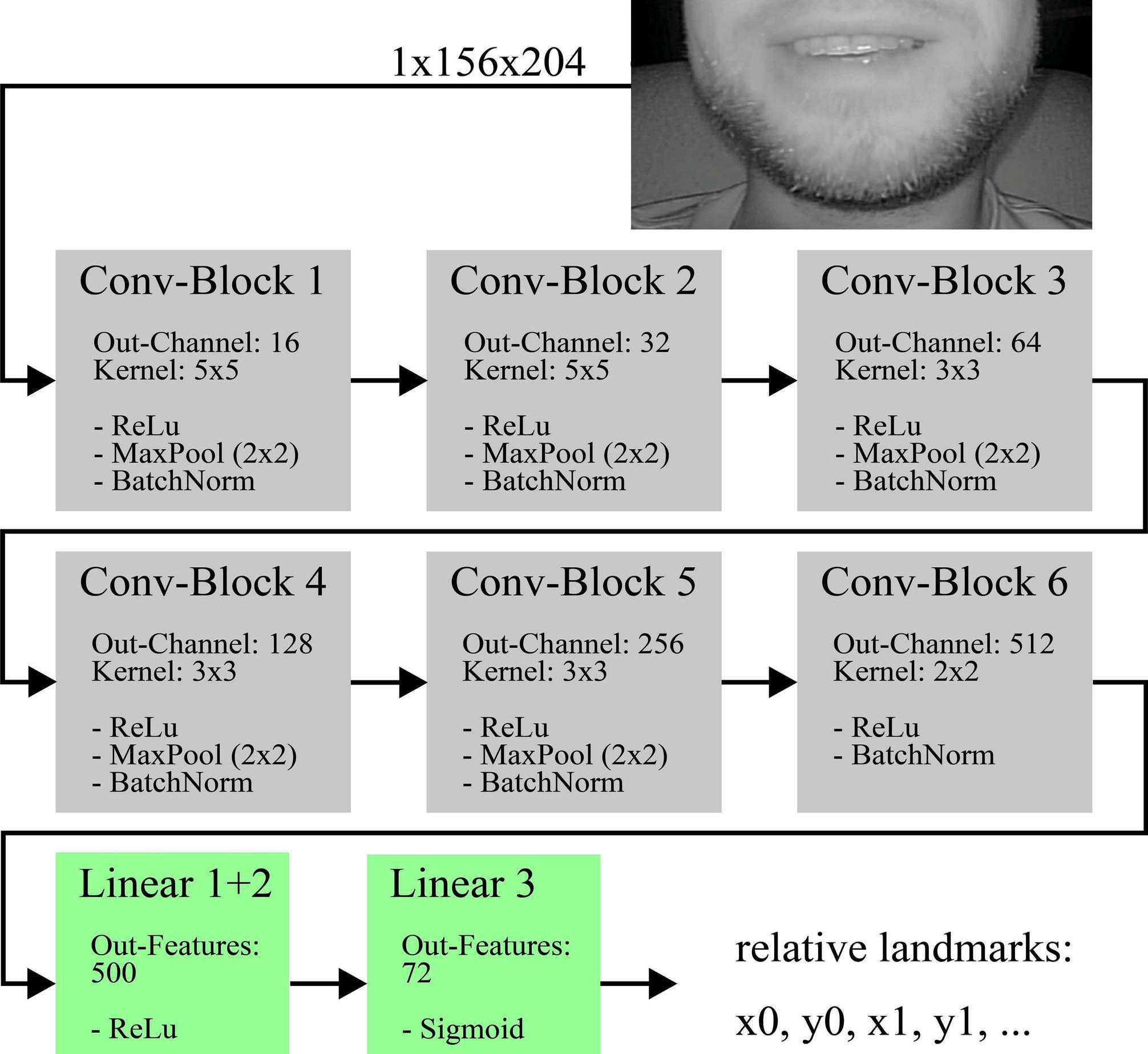}}
\caption{Architecture of the lower-face tracking CNN.}
\label{fig:CNNArchitecture}
\end{figure}

Proven techniques for face tracking fail if the upper face region is covered by an HMD. Therefore, we implemented our own CNN, which is able to extract landmarks from an IR HMC, as shown in Fig\,\ref{fig:Face-Tracking-Vive}. In order to properly train the CNN, labeled training data for the lower face area is required. We generate this labeled training data using the Face Alignment Network (FAN) from Bulat and Tzimiropoulos\,\cite{bulat2017-FAN-HowFarAreWe} to create landmarks for the whole face during training with the helmet mount. We equipped the helmet mount with the same camera at the same angle which is used in the face tracking HMD and we used these images for the training of our CNN. The acquired images were cropped to the lower half of the face and only the remaining landmarks were used as labels for the lower-face tracking CNN, as depicted in Fig.\,\ref{fig:cnnTraning}. Inspired by Lombardi et al.\,\cite{Lombardi18DAM}, the samples are, randomly, flipped horizontally, cropped and rotated to make our method robust to the position of the HMD on the face. We trained the network for over 15 Epochs with a batch size of 8 using the Adam optimizer and a learning rate of 0.001. A data set contains around 10k samples and was divided into training and test data according to the ratio 70:30. As loss function we used the mean squared error. The architecture of the CNN is depicted in Fig. \ref{fig:CNNArchitecture}. 

% lowerFaceRegion-CNN with a data set of landmarks obtained by traditional face tracking method on full face images. We blacked out the RGB data and deleted the landmarks of the upper face region and used it as a labeled training set. The architecture of the CNN is depicted in Fig. X.

% A protruding 3d-printed mount for the HTC Vive Pro Eye was created and equipped with IR LEDs and an IR camera (Fig. X). This way, we are able to obtain facial landmarks of the lower part of the user`s face while wearing an HMD.

\subsection{Eyebrow Tracking}
\begin{figure}[b]
\centerline{\includegraphics[width=\linewidth]{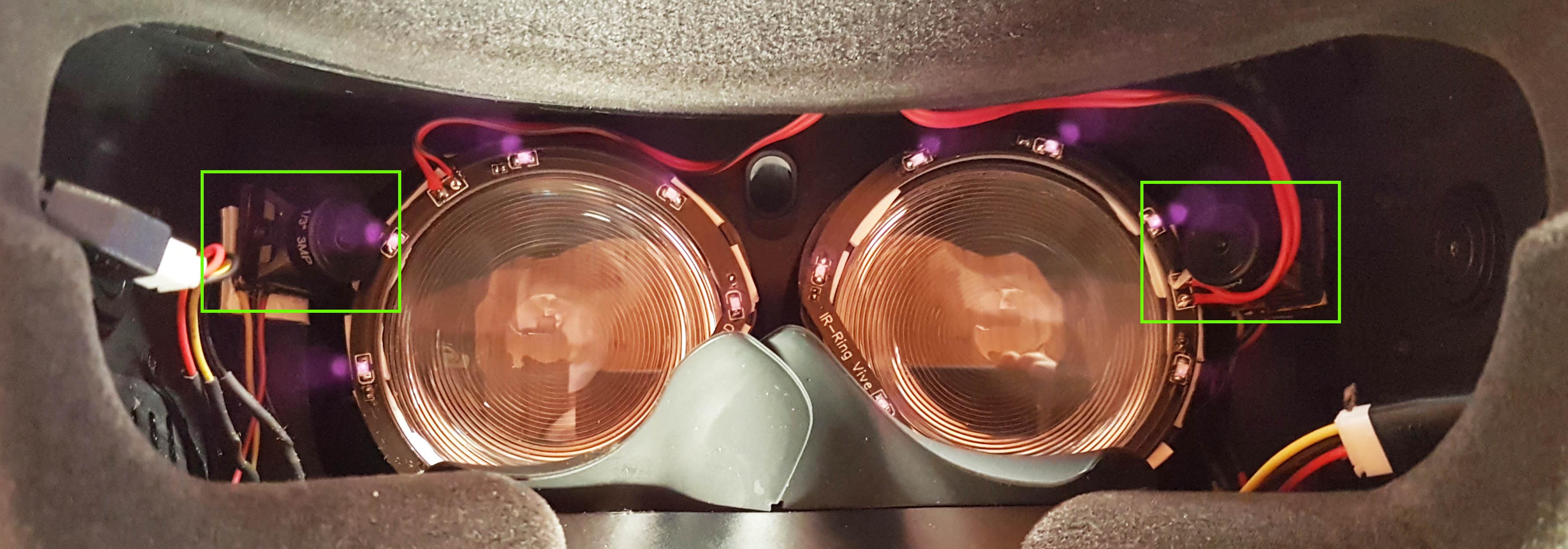}}
\caption{Additional HMCs are attached next the Fresnel lenses (green boxes). We crafted two printed circuit boards (PCBs) with 9 IR LEDs each. They cover the original IR LEDs for eye tracking of the HTC Vive Pro Eye which are located around the Fresnel lenses. This was done to avoid image artifacts in the streams of our HMC for eyebrow tracking.}
\label{fig:insideHMD}
\end{figure}

The SRanipal software does not allow for the collection of tracking data from the position of the eyebrows. As an alternative, using the numerical eye openness values reported by the SRanipal SDK would be one option for predicting them, but usually humans are able to control eye openness and eyebrow position independently, which would only result in an uncertain prediction of the brow position. However, from previous prototypes we have learned that eyebrows convey important non-verbal information cues and can lead to ambiguities of non-verbal communication. Therefore, we attached two HMCs next to the Fresnel lenses of the HMD in order to track them, as shown in Fig.\,\ref{fig:insideHMD}.

The Vive Pro Eye HMD is equipped with 18 IR LEDs around the Fresnel lenses to be able to illuminate the eyes for tracking. The LEDs of this commodity HMD are dimmed with a pulse width modulation and do not allow for synchronizing with our cameras. Without synchronization, our infrared cameras, attached to the inside of the HMD, produce severe image artifacts, which make image processing intractable. Our solution is to cover the original LEDs with our own crafted printed circuit boards (PCB) equipped with IR-LEDs that have the same wavelength. The cost for one PCB ring is around 3\,USD. We adjusted a suitable pulse-width modulation frequency for the LEDs to acquire artifact-free images from the HMCs. A frequency was chosen which had the least impact on eye tracking performance of the Vive.

Moreover, we use OpenCV for tracking the eyebrow position of the user, which allows for the control of 10 facial landmarks in the final FLM. In order to extract information of the eyebrow positions, we turn the HMC streams into an binary image representation. A threshold has to be set to a value that produces images, where the eyebrow is in contrasts to the skin. This way, we can detect the transition between skin and eyebrow in the binary image in order to track the relative position of the eyebrow. Due to the fact that humans have different skin and eyebrow colors, the threshold is individual for each person. 

\subsection{Merging Tracking Data and Reconstruction}
\begin{figure*}[t]
\centerline{\includegraphics[width=\textwidth]{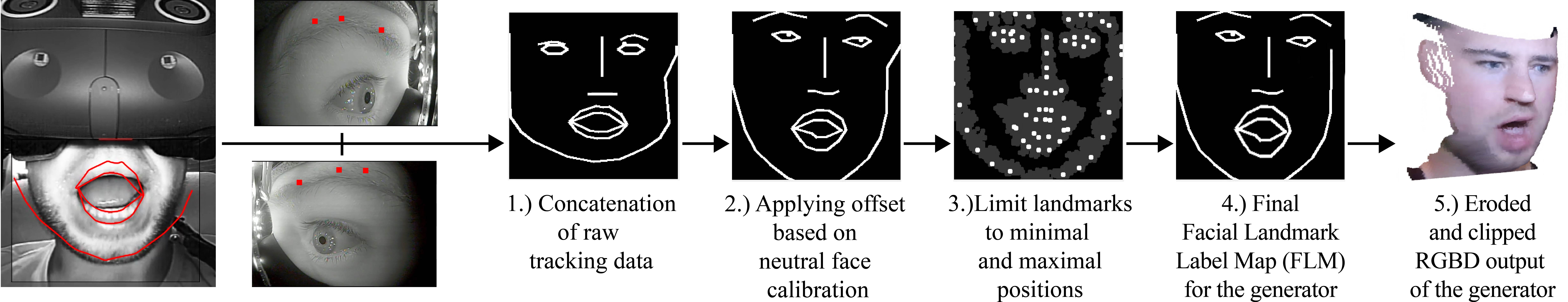}}
\caption{Process from data created by the face tracking HMD to the final FLM and the textured depth map.}
\label{fig:mergeProcess}
\end{figure*}
Each of the 5 HMCs and their corresponding software modules deliver pieces of facial landmark information, which are merged together into the final FLM as input for our GAN. Because of the extreme camera angles the reported landmarks must be modified in 3 steps before they are given to the generator of the RGBD-face-avatar GAN. An overview of these steps is given in Fig.\,\ref{fig:mergeProcess}.

In the first step, each reported landmark of the facial tracking modules is concatenated into an 'uncalibrated' FLM. The second step contains the calibration, which is done by determining the offset between the facial landmarks of a neutral facial expression in the helmet mount (taken during the aforementioned capture procedure for acquiring the training data) and in the face tracking HMD. This offset is continuously added to each of the incoming landmarks. The third step applies a minimum and maximum limit for the x- and y-coordinates of the landmarks. This is important because the quality of the GAN's output decreases if it receives landmarks that were not in the range of the training data set. The fourth step is passing the final FLM to the generator of our GAN in order to reconstruct the RGBD image with the according expression. Finally, we apply an erosion and clipping to the generated depth channel, since the GAN produces randomly sampled points around the face, as can be seen in Fig.\,\ref{fig:Generator} and \ref{fig:Discriminator}.

% Since we do not send image data between the modules, but only a relatively small amount of numbers for the position of the 70 landmarks per label map, the simple OSC protocol is used.  

\section{Results}\label{sec:results}
\begin{figure*}
\centerline{\includegraphics[width=\textwidth]{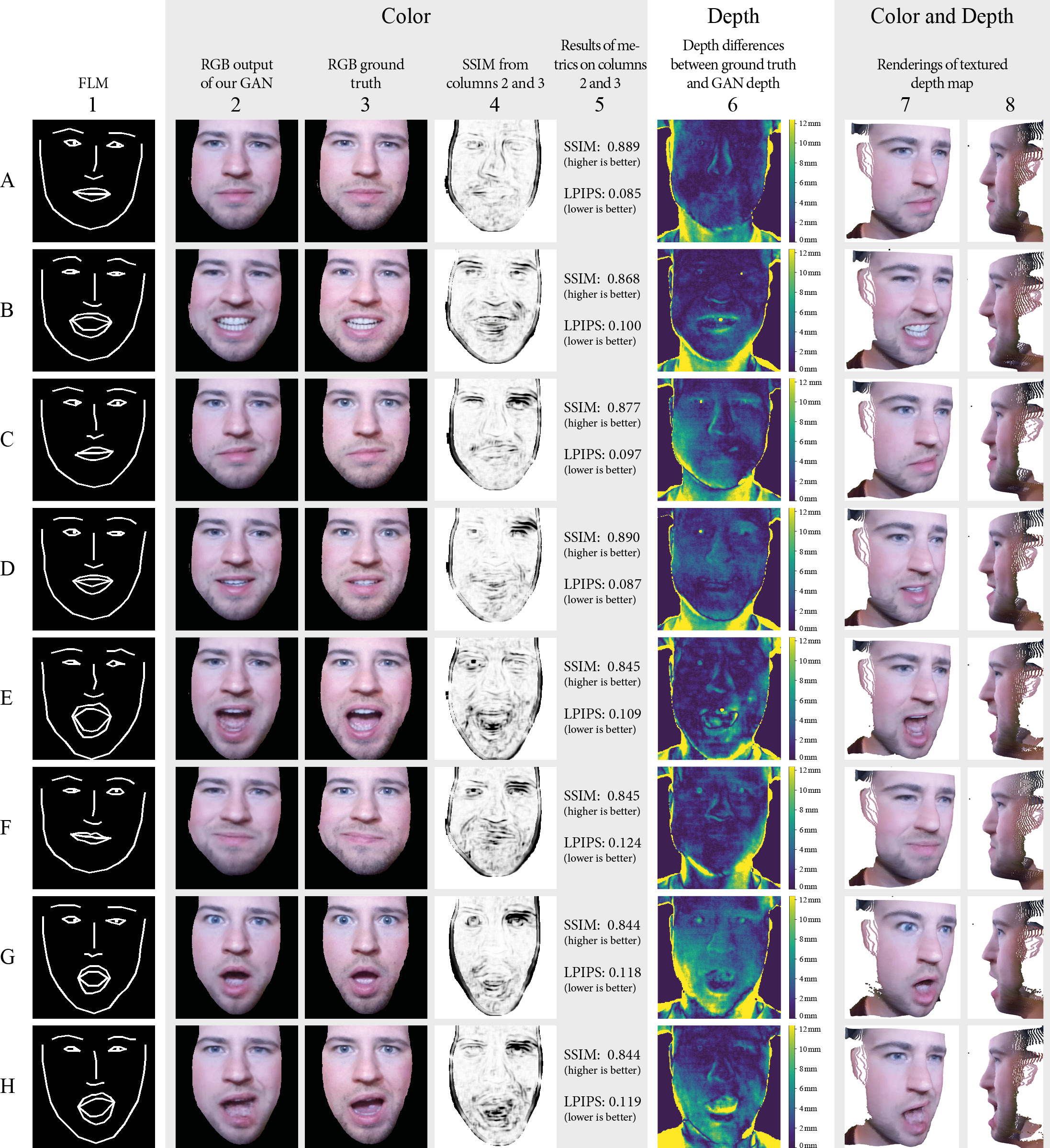}}
\caption{Results: The first column shows the FLM that was given to the generator of our RGBD-face-avatar GAN. Note that these FLMs were not part of the training set and were created by the face tracking HMD. Column 2 shows the generated images by our GAN based on the FLM. Column 3 shows the ground truth expression. They are 'real' images taken by the RGBD camera in the helmet mount. The participant held the expressions while the face tracking HMD was taken off and the helmet mount (Fig.\,\ref{fig:helmet}) was put on and a picture was taken. This procedure produces comparable results since the helmet mount allows for taking an RGBD image in the same way and under the same conditions as the training set was captured. Column 4 shows the difference, based on SSIM\,\cite{ssim}, between the images in columns 2 and 3. The darker the areas are, the greater is the difference. Column 5 shows the quantified results of the metrics SSIM\,\cite{ssim} and LPIPS\,\cite{zhang2018perceptual} (we used version 0.1 from GitHub). Column 6 shows the difference between the ground truth (captured with the helmet mount) and the generated depth. Column 7 and 8 show renderings of the generated textured depth map of our GAN. }
\label{fig:results}
\end{figure*}
% Only minimal offsets of the face were There were small differences in translation and rotation of the face regarding to the image borders which were manually adjusted in order to fit the 

\subsection{Reconstruction quality and identity}
In the following, we present results of our working prototype. Fig.\,\ref{fig:results} shows expressions that were made in the facial tracking HMD. Column 1 shows the resulting FLM generated by the facial tracking HMD and column 2 shows the respective output of our GAN. The participant held the expressions while the face tracking HMD was taken off and the helmet mount (Fig. 1) was put on and a picture was taken, as shown in column 3. This way, we are able to directly compare results with ground truth images. To quantify the difference between the images in column 2 and 3, we used the metrics 'Structural Similarity' (SSIM)\,\cite{ssim} and 'Learned Perceptual Image Patch Similarity' (LPIPS)\,\cite{zhang2018perceptual}. In order to only compare the face without measuring background changes, we determined the facial area based on depth values and rejected all other pixels. 
% The results of the metrics are comparable with a JPEG compression.
% The metrics make a statement 

As Fig.\,\ref{fig:results} shows, the identity of the person can be easily recognized. Compared to the ground truth, the generated images of our GAN have lost a small amount of sharpness and are less detailed, though even small personal characteristics are clearly recognizable in the majority of the images, such as the birthmark on the left cheek at the level between mouth and nose. Most noticeable is the loss of sharpness in the area of the beard. The numerical results of the SSIM metric are comparable to a JPEG compression of a quarter of the original file size of the images in column 3.

The difference of depth values in the face area are mostly below 5\,mm, as shown in column 6. Note that we use the raw depth image of the Kinect and the raw output of our GAN. We do not filter or smooth. In column 7 and 8, we also applied an erosion and clipping to reject the background.

% , as shown in column 5. 'Structural Similarity' (SSIM)\,\cite{ssim} is a perception-based model that considers image degradation as perceived change in structural information. The difference images based on the SSIM between the ground truth and the generated image by our GAN are shown in column 4. Recently, Zhang et al. proposed a new method called 'Perceptual Difference' for quantifying the 'perceptual loss' between image pairs with the aid of CNNs. 
% For a better comparison, the study participant was asked to hold the expression while the face tracking HMD were set off and the helmet mount was put on. This way, it allows for producing comparable results, since the helmet mount allowed for taken an RGBD image in the same way and under the same conditions as the training set was captured.

\subsection{Tracking and quality of expressions}
Our system can recognize and reconstruct a wide variety of expressions. However, humans are highly sensitive to minor discrepancy of facial expressions and small deviations of tracking can lead to conveying different expressions. In row E a slight difference in eyebrow height can be observed, which leads to a rather negative expression than it was actually recorded by our system. Furthermore, the mouth animation is not sufficient for a faithful reconstruction while speaking and inferior to the work of Olszewski et al.\,\cite{Olszewski16HighFid} or Wei et al.\,\cite{Wei19}. We refer the reader to the accompanied videos at the GitHub link for a better comparison in motion. Moreover, expressions which were not part of the training set lead to blurry results and a graceful degradation in this area, as shown in row H around the mouth.

% Moreover, sometimes we observed jitter in gaze and eyebrow tracking, which can lead to an eerie appearance.
\subsection{Scalability to other persons}
\begin{figure*}[t]
\centerline{\includegraphics[width=\linewidth]{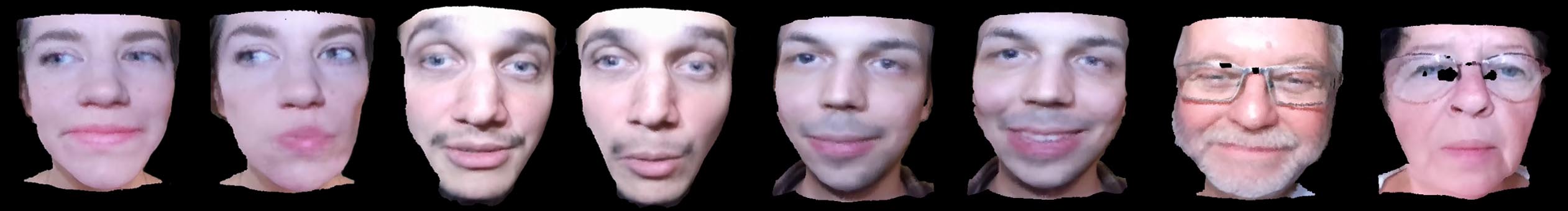}}
\caption{Reconstruction results from the face tracking HMD for 5 persons making different expressions. Our pipeline works regardless of the face shape or personal expression and it does not require any further manual customization to a new person. The images shows different erode and clipping parameters which allows for showing or hiding the neck and upper chest. As a general limitation, we observed that highly reflective surfaces, such as glasses, cannot be reconstructed without failures, as can be seen as black areas in the two images on the far right.}
\label{fig:diffPersonsResults}
\end{figure*}

As Fig\,\ref{fig:diffPersonsResults} shows, we tested our prototype with 5 persons and found that no user specific adaption of our pipeline was necessary. Moreover, each person was able to control his/her avatar with a similar range of expressions. However, we observed that the lower-face tracking CNN does not generalize between persons and only tracks the person who was recorded for the training set as described in section\,\ref{sec:lower-faceTrackingCNN}. 
\subsection{Performance}
After the capture process of the user's face the data set is automatically preprocessed and given to the adversarial training for learning the image-to-image mapping. A data set with 600 samples requires approximately 19\,hours of training on an Nvidia GeForce RTX2080 with PyTorch 1.6 and Python 3.7. For inference and rendering, we use LibTorch 1.6 in combination with a OpenGL-based C++ environment. Generating an RGBD image (8\,bit per channel with 256 by 256 pixels) from a given FLM (8\,bit one channel with 256 by 256 pixels) requires 1.05\,ms. Additional 1.3\,ms are used for stereoscopic rendering of the textured depth map, thus effortlessly enabling refresh rates for the HMD of 90\,fps (equal to 11.1\,ms for a frame). In comparison, the image processing of the face tracking HMD requires more time. Eyebrow tracking, the lower face CNN as well as concatenating the data into the final FLM takes 23\,ms on an Intel i9-9880H and an Nvidia RTX 2080. Nevertheless, due to the limited frame rate of the miniature cameras, the reconstruction rate of the face is bound to 30\,fps (equal to 33.3\,ms for a frame).

\subsection{Network bandwidth}
Once the generator module is transmitted to the remote side (it has always a size of 202MB, irrespective of the shape of the face), the required network bandwidth for driving the avatar is only 67\,kbit/s. This bandwidth is achieved because we do not send raw image data, but only the image positions of 70 landmarks as integers. On the receiver side, the FLM is reconstructed as an image again. Although compression is not the focus of our work, it is interesting to note that the required bandwidth of an RGBD data stream is orders of magnitude higher.

% While inference and rendering is   

% - Grafik wäre sinnvoll
% - Landmark Receive kostet im schnitt: 13.89247691ms und für inference nur 1.058107933ms + Rendern sind 1.3ms + Wenn alles nur auf dem Laptop dann nur noch 18Hz
% - Viele bildverarbeitende Algorithmen
% - 256 x 256px
% - Merger in python with numpy and openCV geschrieben
% - Libtorch 1.6, Visual Studio 2019 compiler

% \subsection{Limitations}
% A limitation of our system is the inability to reconstruct the tongue. 

% Even though our system is able to render teeth while the user is laughing, it often fails to reconstruct the oral cavity without artifacts, as can be seen in rows D and G in Fig.\,\ref{fig:results}. This circumstance complicates a clean reproduction of movements while the user is speaking.

% Another limitation is the reconstruction of expressions that were not part of the training set. For example, row H depicts a 'lip press' expression that was deliberately not trained. Although the generator produces artifacts, the expression can be partly recognized (H2). 

% \subsection{Ablation Study}
% \subsection{Comparison}

\section{Limitations and Future Work}
Although our proposed system proves the feasibility of 3D facial reconstruction in VR, it is one of the early  methods in the field of low-cost solutions and we believe there are exciting opportunities for extending our system in different ways. Since our proposed system is more resource-efficient then expected and the synthesis of an RGBD image takes only 1.05ms at inference, it would be possible to generate multiple views at the same time which could be merged into a single point cloud and would increase the reconstruction quality. Furthermore, the initial capture procedure could be extended by two or more RGBD sensors for recording a person from different views. Another exciting opportunity is to increase the resolution of the FLM and the generated RGBD image with an more advanced network architecture such as Pix2PixHD GAN by Wang et al\,\cite{wang2018pix2pixHD}.

Nevertheless, one of the most important improvements would be a more accurate tracking and an advanced FLM with colors and more landmarks in order to achieve a higher degree of faithful reconstruction of minor changes in expressions. An advanced FLM could also contain information of the tongue, since our current prototype is not able to track it. In addition, we plan to integrate an audio-based approach for further condition our GAN for lip and tongue synchronisation. An interesting audio-based approach was recently presented by Richard et al.\,\cite{richard2020audio}.

\section{Conclusion}
We presented a system that allows for end-to-end capture and reconstruction of facial avatars for telepresence applications in VR using low-cost components. The system is able to preserve individual characteristics of facial expressions and conveys important non-verbal information despite the fact that the user's face is covered by an HMD.

It is the first system for VR telepresence offering personal facial animation that is solely built upon open-source as well as free software and uses commodity hardware. We believe that this work is an important step, demonstrating the feasibility of an open source low-cost system for immersive remote collaboration with the ability to perform an automatic end-to-end capture and reconstruction process of the user's face while requiring no artistic or technical expertise. The latter is an important requirement for a consumer-friendly system and is even more important in times of global challenges such as lowering the energy footprint or pandemics. Furthermore, we are convinced that our system covers only a small range of the potential possibilities offered by the combination of low-cost sensors, commodity hardware and neural networks for immersive telepresence, especially considering the constantly growing processing capabilities of upcoming hardware and the ongoing research in the field of deep artificial networks and neural rendering.
    
\section*{Acknowledgment}
Thanks to the entire MIREVI Group for the great support. This project is sponsored by: German Federal Ministry of Education and Research (BMBF) under the project numbers 16SV8182 and 13FH022IX6. Project names: HIVE-Lab (Health Immersive Virtual Environment Lab) and Interactive body-near production technology 4.0 (german: 'Interaktive körpernahe Produktionstechnik 4.0' (iKPT4.0)).
% Dummy Acknowledgment

% Notes:\\
% - markerless tracking \\
% - compression - much less bandwidth, of once submitted -> generator can be send over network and remote app receives only 140 shorts -> super lightweight! Ist ein vorteil gegenüber rein video basierten systemen \\
% - if accepted  \\

%     - Grafiken werden besser \\
    
%     - Evaluierung wird überarbeitet (bessere diff bilder und gleiches t shirt)  \\
% 

\bibliographystyle{IEEEtran}
\bibliography{ladwig}

\begin{thebibliography}{10}
\providecommand{\url}[1]{#1}
\csname url@samestyle\endcsname
\providecommand{\newblock}{\relax}
\providecommand{\bibinfo}[2]{#2}
\providecommand{\BIBentrySTDinterwordspacing}{\spaceskip=0pt\relax}
\providecommand{\BIBentryALTinterwordstretchfactor}{4}
\providecommand{\BIBentryALTinterwordspacing}{\spaceskip=\fontdimen2\font plus
\BIBentryALTinterwordstretchfactor\fontdimen3\font minus
  \fontdimen4\font\relax}
\providecommand{\BIBforeignlanguage}[2]{{%
\expandafter\ifx\csname l@#1\endcsname\relax
\typeout{** WARNING: IEEEtran.bst: No hyphenation pattern has been}%
\typeout{** loaded for the language `#1'. Using the pattern for}%
\typeout{** the default language instead.}%
\else
\language=\csname l@#1\endcsname
\fi
#2}}
\providecommand{\BIBdecl}{\relax}
\BIBdecl

\bibitem{UncannyEffectAusFidler07}
J.~{Seyama} and R.~S. {Nagayama}, ``{The Uncanny Valley: Effect of Realism on
  the Impression of Artificial Human Faces},'' \emph{Presence}, vol.~16, no.~4,
  2007.

\bibitem{Tewari2020NeuralSTAR}
A.~Tewari, O.~Fried, J.~Thies, V.~Sitzmann, S.~Lombardi, K.~Sunkavalli,
  R.~Martin-Brualla, T.~Simon, J.~Saragih, M.~Nie{\ss}ner, R.~Pandey,
  S.~Fanello, G.~Wetzstein, J.-Y. Zhu, C.~Theobalt, M.~Agrawala, E.~Shechtman,
  D.~B. Goldman, and M.~Zollh{\"o}fer, ``{State of the Art on Neural
  Rendering},'' \emph{Computer Graphics Forum}, 2020.

\bibitem{STAR-3DMM}
B.~Egger, W.~A.~P. Smith, A.~Tewari, S.~Wuhrer, M.~Zollhoefer, T.~Beeler,
  F.~Bernard, T.~Bolkart, A.~Kortylewski, S.~Romdhani, C.~Theobalt, V.~Blanz,
  and T.~Vetter, ``{3D Morphable Face Models—Past, Present, and Future},''
  \emph{ACM Trans. Graph.}, vol.~39, no.~5, Jun. 2020.

\bibitem{Lombardi18DAM}
S.~Lombardi, J.~Saragih, T.~Simon, and Y.~Sheikh, ``{Deep Appearance Models for
  Face Rendering},'' \emph{ACM Trans. Graph.}, vol.~37, no.~4, Jul. 2018.

\bibitem{Wei19}
S.-E. Wei, J.~Saragih, T.~Simon, A.~W. Harley, S.~Lombardi, M.~Perdoch,
  A.~Hypes, D.~Wang, H.~Badino, and Y.~Sheikh, ``{VR Facial Animation via
  Multiview Image Translation},'' \emph{ACM Trans. Graph.}, vol.~38, no.~4,
  Jul. 2019.

\bibitem{ModularCodecAvatars20}
H.~Chu, S.~Ma, F.~De~la Torre, S.~Fidler, and Y.~Sheikh, ``Expressive
  telepresence via modular codec avatars,'' in \emph{European Conference on
  Computer Vision - ECCV}, 2020.

\bibitem{richard2020audio}
A.~Richard, C.~Lea, S.~Ma, J.~Gall, F.~de~la Torre, and Y.~Sheikh, ``Audio- and
  gaze-driven facial animation of codec avatars,'' 2020, only preprint
  available, no conference.

\bibitem{starMonoFace18}
M.~Zollh{\"o}fer, J.~Thies, P.~Garrido, D.~Bradley, T.~Beeler, P.~P{\'e}rez,
  M.~Stamminger, M.~Nie{\ss}ner, and C.~Theobalt, ``{State of the Art on
  Monocular 3D Face Reconstruction, Tracking, and Applications},''
  \emph{Computer Graphics Forum}, vol.~37, 2018.

\bibitem{Li15FacialHMD}
H.~Li, L.~Trutoiu, K.~Olszewski, L.~Wei, T.~Trutna, P.-L. Hsieh, A.~Nicholls,
  and C.~Ma, ``{Facial Performance Sensing Head-Mounted Display},'' \emph{ACM
  Trans. Graph.}, vol.~34, no.~4, Jul. 2015.

\bibitem{Casas16Rapid}
D.~Casas, A.~Feng, O.~Alexander, G.~Fyffe, P.~Debevec, R.~Ichikari, H.~Li,
  K.~Olszewski, E.~Suma, and A.~Shapiro, ``{Rapid Photorealistic Blendshape
  Modeling from RGB-D Sensors},'' in \emph{Computer Animation and Social
  Agents}, ser. CASA, 2016.

\bibitem{Olszewski16HighFid}
K.~Olszewski, J.~J. Lim, S.~Saito, and H.~Li, ``{High-Fidelity Facial and
  Speech Animation for VR HMDs},'' \emph{ACM Trans. Graph.}, vol.~35, no.~6,
  Nov. 2016.

\bibitem{Thies18FaceVR}
J.~Thies, M.~Zollh\"{o}fer, M.~Stamminger, C.~Theobalt, and M.~Nie\ss{}ner,
  ``{FaceVR: Real-Time Gaze-Aware Facial Reenactment in Virtual Reality},''
  \emph{ACM Trans. Graph.}, vol.~37, no.~2, Jun. 2018.

\bibitem{3dmm}
V.~Blanz and T.~Vetter, ``{A Morphable Model for the Synthesis of 3D Faces},''
  in \emph{Proceedings of the 26th Annual Conference on Computer Graphics and
  Interactive Techniques}, ser. SIGGRAPH, 1999.

\bibitem{UnpairedI2I-Zhu17}
J.~{Zhu}, T.~{Park}, P.~{Isola}, and A.~A. {Efros}, ``{Unpaired Image-to-Image
  Translation Using Cycle-Consistent Adversarial Networks},'' in \emph{2017
  IEEE International Conference on Computer Vision (ICCV)}, 2017.

\bibitem{bulat2017-FAN-HowFarAreWe}
A.~Bulat and G.~Tzimiropoulos, ``{How far are we from solving the 2D \& 3D Face
  Alignment problem? (and a dataset of 230,000 3D facial landmarks)},'' in
  \emph{International Conference on Computer Vision (ICCV)}, 2017.

\bibitem{Ir-EyeGaze-Works}
X.~Xiong, Z.~Liu, Q.~Cai, and Z.~Zhang, ``{Eye Gaze Tracking Using an RGBD
  Camera: A Comparison with a RGB Solution},'' ser. UbiComp '14 Adjunct, 2014.

\bibitem{pix2pix2016}
P.~Isola, J.-Y. Zhu, T.~Zhou, and A.~A. Efros, ``{Image-to-Image Translation
  with Conditional Adversarial Networks},'' \emph{IEEE Conference on Computer
  Vision and Pattern Recognition (CVPR)}, 2017.

\bibitem{ViveEye}
``{HTC Vive Eye Pro},''
  \url{https://www.vive.com/eu/product/vive-pro-eye/overview/}, 2020, accessed:
  2020-08-28.

\bibitem{SRanipal}
``{VIVE Eye Tracking SDK (SRanipal)},''
  \url{https://developer.vive.com/resources/vive-sense/sdk/vive-eye-tracking-sdk-sranipal/},
  2020, accessed: 2020-09-03.

\bibitem{ssim}
{Zhou Wang}, A.~C. {Bovik}, H.~R. {Sheikh}, and E.~P. {Simoncelli}, ``Image
  quality assessment: from error visibility to structural similarity,''
  \emph{IEEE Transactions on Image Processing}, vol.~13, no.~4, 2004.

\bibitem{zhang2018perceptual}
R.~Zhang, P.~Isola, A.~A. Efros, E.~Shechtman, and O.~Wang, ``{The Unreasonable
  Effectiveness of Deep Features as a Perceptual Metric},'' in \emph{IEEE
  Conference on Computer Vision and Pattern Recognition (CVPR)}, 2018.

\bibitem{wang2018pix2pixHD}
T.-C. Wang, M.-Y. Liu, J.-Y. Zhu, A.~Tao, J.~Kautz, and B.~Catanzaro,
  ``{High-Resolution Image Synthesis and Semantic Manipulation with Conditional
  GANs},'' in \emph{IEEE Conference on Computer Vision and Pattern Recognition
  (CVPR)}, 2018.

\end{thebibliography}
\end{document}